\documentstyle[12pt]{article}

\begin{document}
\begin{titlepage}
\begin{flushright}
solv-int/0102025
\end{flushright}
\vskip.3in
\begin{center}
{\Large \bf q-deformed supersymmetric $t$-$J$ model with a boundary}
\vskip.3in
{\large
Bo-Yu Hou $^{a}$
\footnote{E-mail: byhou@phy.nwu.edu.cn},~
 Wen-Li Yang $^{a,b}$
\footnote{E-mail: wlyang@phy.nwu.edu.cn},~
 Yao-Zhong Zhang $^{c}$ \footnote{E-mail: yzz@maths.uq.edu.au}
and
 ~Yi Zhen $^{a}$
\footnote{E-mail: zheny@phy.nwu.edu.cn}}
\vskip.2in
{\em
$~^a$ Institute of Modern Physics, Northwest University,
Xian 710069, China \\
$~^b$ Physikalisches Institut der Universitat Bonn, Nussallee 12,
53115
Bonn, Germany\\
$~^c$ Department of Mathematics, University of Queensland, Brisbane,
Qld 4072, Australia}
\end{center}

\vskip 2cm
\begin{center}
{\bf Abstract}
\end{center}
The q-deformed supersymmetric $t$-$J$ model on a semi-infinite 
lattice is diagonalized
by using the level-one vertex operators of the	quantum affine superalgebra
$U_q[\widehat{sl(2|1)}]$. We give the bosonization
of the boundary states.
We give an integral expression of the correlation
functions of the boundary model,
and derive the difference equations which they satisfy.

\vskip 2cm

\end{titlepage}

\def\a{\alpha}
\def\b{\beta}
\def\d{\delta}
\def\e{\epsilon}
\def\ve{\varepsilon}
\def\g{\gamma}
\def\k{\kappa}
\def\l{\lambda}
\def\o{\omega}
\def\p{\phi}
\def\t{\theta}
\def\s{\sigma}
\def\u{\mu}
\def\D{\Delta}
\def\L{\Lambda}
\def\T{{\cal T}}

\def\R{\overline{R}}
\def\S{{sl(N|1)}}
\def\hS{{\widehat{sl(N|1)}}}
\def\hG{{\widehat{gl(N|N)}}}
\def\R{{\cal R}}
\def\hR{{\hat{\cal R}}}
\def\C{{\bf C}}
\def\P{\Phi}
\def\Z2{{{\bf Z}_2}}
\def\Z{{\bf Z}}
\def\T{{\cal T}}
\def\F{{\cal F}}
\def\V{\overline{V}}
\def\trho{{\tilde{\rho}}}
\def\tphi{{\tilde{\phi}}}
\def\tT{{\tilde{\cal T}}}
\def\ta{{U_q[A^{(2)}_2]}}
\def\uqsnh{{U_q[\widehat{sl(N|1)}]}}
\def\uq2h{{U_q[\widehat{sl(2)}]}}
\def\uq1h{{U_q[\widehat{gl(1|1)}]}}
\def\uqs2h{{U_q[\widehat{sl(2|1)}]}}


\def\beq{\begin{equation}}
\def\eeq{\end{equation}}
\def\bea{\begin{eqnarray}}
\def\eea{\end{eqnarray}}
\def\ba{\begin{array}}
\def\ea{\end{array}}
\def\no{\nonumber}
\def\lt{\left}
\def\rt{\right}
\newcommand{\bq}{\begin{quote}}
\newcommand{\eq}{\end{quote}}

\newtheorem{Theorem}{Theorem}
\newtheorem{Definition}{Definition}
\newtheorem{Proposition}{Proposition}
\newtheorem{Lemma}{Lemma}
\newtheorem{Corollary}{Corollary}
\newcommand{\proof}[1]{{\bf Proof. }
	#1\begin{flushright}$\Box$\end{flushright}}

\newcommand{\sect}[1]{\setcounter{equation}{0}\section{#1}}
\renewcommand{\theequation}{\thesection.\arabic{equation}}
\sect{Introduction}

Integrable models with quantum superalgebra symmetries have
been the focus of recent studies \cite{Ess92, Foe93,Bra95,Pfa96,Ram96,Fan00} 
in the context of
strongly correlated fermion systems, a subject of
high-profile international research activity because of
their relevance to high-$T_c$ superconductivity. The
investigations to these models have largely been carried out within the
framework of QISM and Bethe ansatz method. The exception  are
the works in \cite{Yan99}, where the algebraic analysis method,
developed in \cite{Dav93,Jim94} and generalized in
\cite{Idz94,Koy94,Luk96,Hon98,Hou99}, was used to diagonalize the
supersymmetric $t$-$J$ model and its multi-component version directly on an
infinite lattice. 

The algebraic analysis method \cite{Dav93,Jim94}, which we will call the
vertex operator method, was formulated with the help of the
level-one q-vertex operators \cite{Fre92} and highest weight
representations of quantum affine algebras. 
The vertex operator method 
was later extended in \cite{Jim95} to treat
integrable models with boundary interactions \cite{Sky88,Gho94}.
It was shown in \cite{Jim95} how the space
of states of the boundary XXZ spin-$\frac{1}{2}$ chain on a semi-infinite
lattice can be described in terms
of level-one q-vertex operators of $U_q(\widehat{sl_2})$, and how the
correlation functions can be computed by the vertex operators.
Several other models have been analysed by means of this approach
\cite{Miw97,Fur00,Yan01,Hou97}.

In this paper, we study the q-deformed supersymmetric  $t$-$J$
model with an integrable boundary. We will work directly on a 
semi-infinite lattice. As is known, the
q-deformed supersymmetric  $t$-$J$ model on an infinite lattice
(i.e. without a boundary) has as
its symmetry algebra the  quantum affine superalgebra $\uqs2h$ \cite{Yan99}.
On a finite lattice with diagonal boundary reflection K-matrices
this model was  solved in \cite{Fan00} by the Bethe ansatz method.
Here we adapt the vertex operator method.
We will diagonalize the	boundary model Hamiltonian directly on the semi-infinite
lattice, and moreover compute the correlation
functions of the boundary model.

This paper is organized as follows. In section 2, we describe the
vertex operator approach to the q-deformed supersymmetric  $t$-$J$
model on the semi-infinite lattice.
In section 3, we study the bosonic
realization of the boundary states associated with the level-one
highest weight representation of $\uqs2h$.
In section 4, we compute the correlation functions of
the local operators (including	the spin operator $S^z_1$ ) and derive the
difference equations which they satisfy.
In appendix A, we review the bosonization of 
$\uqs2h$ at level-one and the associated vertex operators. 

\sect{Boundary q-deformed supersymmetric  $t$-$J$ model}

\subsection{q-deformed supersymmetric $t$-$J$ model on a finite lattice}

In this section, we recall some facts about the 
q-deformed supersymmetric $t$-$J$ model on a finite lattice.
Throughout this paper, we fix $q$ such that $|q|<1$.


Let $V$ be the 3-dimensional graded vector space  and
$E_{ij}$ be the $3\times 3$ matrix whose $(i,j)$-element is
unity and zero otherwise. The grading of the basis vectors
$v_1$, $v_2$, $v_3$ of $V$ is chosen to be $[v_1]=[v_2]=1$,
$[v_3]=0$. Let $V^*$ be the dual space and
$\{v^*_1,v^*_0,v^*_{-1}\}$  the dual basis vectors.
Denote by $V_z$ (resp. $V^{*S}_z$)  the 3-dimensional level-0
representation	(resp. dual representation) of
$\uqs2h$ associated with $V$.  Let $R(z)\in
End(V\otimes V)$
be the R-matrix of $\uqs2h$ with matrix elements defined by
\bea
\label{R}
R(z)(v_i\otimes v_j)=\sum_{k,l}R^{ij}_{kl}(z)v_k\otimes v_l, \ \ \ \
 \forall v_i, v_j, v_k, v_l\in V,\no
\eea
where 
\begin{eqnarray*}
& &R^{33}_{33}(\frac{z_1}{z_2})=-\frac{z_1q^{-1}-z_2q}{z_1q-z_2q^{-1}},\ \
R^{23}_{23}(\frac{z_1}{z_2})=-\frac{z_1-z_2}{z_1q-z_2q^{-1}},\ \
R^{32}_{23}(\frac{z_1}{z_2})=\frac{(q-q^{-1})z_2}{z_1q-z_2q^{-1}},\\
& &R^{32}_{32}(\frac{z_1}{z_2})=-\frac{z_1-z_2}{z_1q-z_2q^{-1}},\ \
R^{23}_{32}(\frac{z_1}{z_2})=\frac{(q-q^{-1})z_1}{z_1q-z_2q^{-1}},\ \
R^{22}_{22}(\frac{z_1}{z_2})=-1,\\
& &R^{13}_{13}(\frac{z_1}{z_2})=-\frac{z_1-z_2}{z_1q-z_2q^{-1}},\ \
R^{31}_{13}(\frac{z_1}{z_2})=\frac{(q-q^{-1})z_2}{z_1q-z_2q^{-1}},\ \
R^{31}_{31}(\frac{z_1}{z_2})=-\frac{z_1-z_2}{z_1q-z_2q^{-1}},\\
& &R^{13}_{31}(\frac{z_1}{z_2})=\frac{(q-q^{-1})z_1}{z_1q-z_2q^{-1}},\ \
R^{12}_{12}(\frac{z_1}{z_2})=-\frac{z_1-z_2}{z_1q-z_2q^{-1}},\ \
R^{21}_{12}(\frac{z_1}{z_2})=-\frac{(q-q^{-1})z_2}{z_1q-z_2q^{-1}},\\
& &R^{21}_{21}(\frac{z_1}{z_2})=-\frac{z_1-z_2}{z_1q-z_2q^{-1}},\ \
R^{12}_{21}(\frac{z_1}{z_2})=-\frac{(q-q^{-1})z_1}{z_1q-z_2q^{-1}},\ \
R^{11}_{11}(\frac{z_1}{z_2})=-1 ,\\
& &R^{ij}_{kl}=0\ \ ,\ \ {\rm otherwise}.
\end{eqnarray*}
The R-matrix satisfies	the graded Yang-Baxter equation(YBE) on
$V\otimes V\otimes V$
\begin{eqnarray*}
R_{12}(z)R_{13}(zw)R_{23}(w)=R_{23}(w)R_{13}(zw)R_{12}(z),
\end{eqnarray*}
and moreover enjoys: (i) initial condition,
$R(1)=P$ with $P$ being
the graded permutation operator; (ii) unitarity condition,
$R_{12}(\frac{z}{w})R_{21}(\frac{w}{z})=1$, where
$R_{21}(z)=PR_{12}(z)P$; and (iii) crossing-unitarity,
\begin{eqnarray*}
R^{-1,st_1}(z)\lt((M\otimes 1)R(zq^{-2})
(M\otimes 1)\rt)^{st_1}=1\otimes 1 ,
\end{eqnarray*}
where
\begin{eqnarray}
\label{CU}
M\equiv q^{2\overline{\rho}}\stackrel{def}{=}
\left(
   \begin{array}{ccc}
     q^{2\rho_1}&&\\&q^{2\rho_2}&\\&&q^{2\rho_3}
    \end{array}
  \right)
=\left(
    \begin{array}{ccc}
     1&&\\&q^{-2}&\\&&q^{-2}
    \end{array}
  \right) .
\end{eqnarray}
The various supertranspositions of the R-matrix are given by
\begin{eqnarray*}
& &(R^{st_1}(z))^{kl}_{ij}=R(z)^{il}_{kj}(-1)^{[i]([i]+[k])},\ \
(R^{st_2}(z))^{kl}_{ij}=R(z)^{kj}_{il}(-1)^{[j]([j]+[l])},\\
& &(R^{st_{12}}(z))^{kl}_{ij}=
R(z)^{ij}_{kl}(-1)^{([i]+[j])([i]+[j]+[l]+[k])}
=R(z)^{ij}_{kl} .
\end{eqnarray*}

Following Sklyanin \cite{Sky88}, we construct the transfer
matrix of an integrable finite chain, with an open boundary condition
described by a reflection K-matrix $K(z)$. 
Here $K(z)$ is a solution of
the graded reflection equation
\bea
K_2(z_2)R_{21}(z_1z_2)K_1(z_1)R_{12}(z_1/z_2)=R_{21}(z_1/z_2)
K_1(z_1)R_{12}(z_1z_2)K_2(z_2).\label{Ref}
\eea
With
appropriate normalization, we can show that this $K(z)$ obeys the relations
\bea
\begin{array}{ll}
K(1)=1,&
({\rm Boundary ~initial~condition}),\\
K(z)K(z^{-1})=1,&({\rm Boundary~ unitarity}),\\
\overline{K}(z)\overline{K}(z^{-1})=1,&
({\rm Boundary~ crossing-unitarity}),
\end{array}\label{bc}
\eea
where $\overline{K}(z)$ is defined by
\bea
\overline{K}(z)=-\sum_{\a,\b}R(z^2)^{\a
j}_{i\b}(-1)^{[i]+[j]+[j][\b]+[\a][\b]}K^{\b}_{\a}(z^{-1}q^{-1})q^{2\rho_{\a}}.
\label{bc1}
\eea
The third relation is 
the graded extension of the boundary crossing-unitarity
\cite{Gho94,Jim95,Doi99}.

The transfer matrix of the q-deformed supersymmetric $t$-$J$ model on a
finite chain with the open
boundary condition is constructed from $R(z)$ and $K(z)$ via \cite{Sky88,Bra98}
\bea
T^{{\rm fin}}_{B}(z)=str_{V_0}(K^+(z)\T(z^{-1})K(z)\T(z)),\label{tf}
\eea
where $K^+(z)=K(-z^{-1}q^{-3})^{st}M$ and
\bea
\T(z)=R_{01}(z)\cdots R_{0N}(z)\in {End}(V_0\otimes V_1\otimes \cdots
\otimes V_N)\no
\eea
is the double-row monodromy matrix. The supertrace is
defined as $str(A)=\sum (-1)^{[i]}A_{ii}$.

It can be verified that $T^{{\rm fin}}_{B}(z)$ form a commuting family,
 $[T^{{\rm fin}}_{B}(z),T^{{\rm fin}}_{B}(w)]=0$. The Hamiltonian
of the boundary q-deformed supersymmetric $t$-$J$ 
model is given by \cite{Sky88,Fan00}
\bea
H_B^{{\rm fin}}=\frac{d}{dz}T^{{\rm
fin}}_B(z)|_{z=1}=\sum_{j=1}^{N-1}h_{j,j+1} +
\frac{1}{2}\frac{d}{dz}K(z)|_{z=1}+\frac{str_{V_0}(K^+(1)h_{0,N})}
{K^+(1)},\label{hf}
\eea
where $h_{j,j+1}=P_{j,j+1}\frac{d}{dz}R_{j,j+1}(z)|_{z=1}$.

The transfer matrix (\ref{tf}) with diagonal reflection K-matrices was
diagonalized by the  Bethe ansatz method in \cite{Fan00}.

\subsection{q-deformed supersymmetric $t$-$J$ model on a semi-infinite lattice}

In this paper, we restrict ourselves to the
diagonal reflection K-matrix of the form
\bea
K(z)=f(z)\lt(\begin{array}{ccc}
\frac{1-rz}{z-r}z&0&0\\
0&1&0\\
0&0&1
\end{array}
\rt)
,~~f(z)=\frac{\p(z|r)}{\p(z^{-1}|r)},~~
\p(z|r)=\frac{1}{1-rz},
\eea
where $r$ is an arbitrary parameter which is related with
the boundary interaction\cite{Gho94,Jim95}.
One can check that such a K-matrix satisfies the boundary unitarity and
crossing-unitarity (\ref{bc}).

We now consider Hamiltonian (\ref{hf})
in the semi-infinite limit:
\bea
H^{{\rm fin}}_B|_{N\longrightarrow \infty}=\sum_{j=1}^{\infty}
h_{j,j+1} +\D,
\eea
where $\D=\frac{1}{2}\frac{d}{dz}f(z)|_{z=1}$ acts formally
on the left-infinite tensor product space
\bea
\cdots \otimes V\otimes V.\label{lf}
\eea

As mentioned in the introduction, the q-deformed supersymmetric $t$-$J$ model
on an infinite lattice has $\uqs2h$ as its symmetry algebra.
Let $V(\u_{\a})$ be the level-one
irreducible highest weight $\uqs2h$-modules with highest weight
$\u_{\a}$,  $\a\in {\bf Z}$ (see (\ref{Wei}) and \cite{Yan99}).
Consider the level-one vertex operators which are intertwining operators
between $V(\u_{\a})$ and $V(\u_{\b})$. It has been
shown in \cite{Yan99} that the following type I vertex
operators $\Phi(z)$ exist, 
$\Phi^*(z)$ which interwine the level-one irreducible highest
weight $\uqs2h$-modules $V(\u_{\a})$
\bea
\Phi(z):~V(\u_{\a})\longrightarrow V(\u_{\a-1})\otimes V_z,
~~\Phi^*(z):~V(\u_{\a})\longrightarrow V(\u_{\a+1})\otimes
V^{*S}_z.
\eea
(See Appendix A for more details about $V(\u_{\a})$ and its associated
vertex operators.)
Therefore following \cite{Dav93,Yan99,Jim95,Yan01},
we can write
the transfer matrix of the q-deformed supersymmetric $t$-$J$ model on
the semi-infinite lattice as
\bea
T_B(z)&=&-\sum_{i,j=1}^{3}\Phi^{*}_i(z^{-1})K^j_i(z)\Phi_j(z)(-1)^{[i]}
\no\\
&=&\sum_{i,j=1}^{3}q^{-2\rho_j}\Phi_j(z)\overline{K}^j_i(z^{-1}q^{-1})
\Phi^*_i(z^{-1}),\label{TF}
\eea
where $\P_i(z)$ and $\P^*_j(z)$ are the components of the  $\uqs2h$ vertex
operators of type I (see (\ref{VoxI})).
We have used the exchange relations of
vertex operators (\ref{ZF3}) and the definition of $\overline{K}(z)$
(\ref{bc1}) in the above equation.

We remark that the transfer matrix $T_B(z)$ given by (\ref{TF}) is an operator
with the property
\bea
T(z):~~V(\u_{\a})\longrightarrow V(\u_{\a}),~~ \a\in Z.\no
\eea
The commutativity of the transfer matrix (\ref{TF}),
$[T_B(z),T_B(w)]=0$ , then follows from (\ref{ZF1}) and (\ref{Ref}).
Moreover by (\ref{ZF1}), (\ref{DEF1}) and (\ref{DEF2}), one can show
\bea
&&T_B(1)=id, ~~~~T_B(z)T_B(z^{-1})=id,\\
&&T_B(z)T_B(z^{-1}q^{-2})=id.
\eea
These relations correspond to the boundary initial condition,
boundary unitarity and boundary crossing-unitarity (\ref{bc}) of the
K-matrix, respectively. In terms of the transfer matrix, the
q-deformed supersymmetric $t$-$J$ model Hamiltonian on the
semi-infinite lattice is given
by
\bea
H=\frac{d}{dz}T_B(z)|_{z=1}.\label{Ham}
\eea

Following \cite{Yan99},	 we define the local
operators acting on the n-th site:
\bea
&&E^{(1)}_{i,j}=-\P^*_i(1)\P_j(1)(-1)^{[j]},\\
&&E^{(n)}_{i,j}=\sum_{m}(-1)^{([i]+[j])[m]+[m]}\P^*_m(1)E^{(n-1)}_{i,j}\P_m(1),
~~n=2,3,\cdots.
\eea
In particular, we have the spin operator $S^z_1$
\bea
S^z_1=\frac{1}{2}(E^{(1)}_{11}-E^{(1)}_{22})=\frac{1}{2}
\lt.\{\P^*_1(1)\P_1(1)-\P^*_{2}(1)\P_{2}(1)\rt.\}.\no
\eea

\section{The boundary states}
In this section we construct the bosonic
boundary state $|\a;r>_B$ and its dual state ${}_B<r;\a|$, which satisfy
\bea
T_B(z)|\a;r>_B=|\a;r>_B,~~{}_B<r;\a|T_B(z)={}_B<r;\a|.\label{Vac}
\eea
By (\ref{DEF1}) and
(\ref{TF}), the above eigenvalue problem is
equivalent to
\bea
\P_i(z^{-1})|\a;r>_B&=&\sum_{j}K^j_i(z)\P_j(z)|\a;r>_B,\label{Bst1}\\
{}_B<r;\a|\P^*_j(z)(-1)^{[j]}&=&\sum_{i}{}_B<r;\a|\P^*_i(z^{-1})K^j_i(z)
(-1)^{[i]}.\label{Bst2}
\eea

\subsection{The boundary state in $V(\L_0)$}

Firstly, we consider the boundary state $|0>_B\in
V(\u_0)$ (or $V(\L_0)$). As is shown in Appendix A, 
$V(\u_0)=\eta_0\xi_0F_{(0;\b)}$
and the highest weight vector	$|\L_0>=|\b,\b,\b,0>$  satisfies
\bea
\eta_0|\L_0>=0.\no
\eea
So we make the following ansatz \cite{Jim95}:
\bea
|0;r>_{B} &=&e^{F_0(r)}|\L_0>, \\
F_0(r) &=&\frac{1}{2}\sum_{m=1}^{\infty }\frac{m}{\lbrack m\rbrack
^{2}}\alpha _{m}\{h_{-m}^{1}h_{-m}^{\ast 1}+h_{-m}^{2}h_{-m}^{\ast
2}+c_{-m}c_{-m}\}+\sum_{m=1}^{\infty }\{\beta _{m}^{1}h_{-m}^{1}\no \\
&&+\beta _{m}^{2}h_{-m}^{2}+\beta _{m}^{3}c_{-m}\},
\eea
where $\alpha _{m},\beta _{m}^{1},\beta _{m}^{2},\beta
_{m}^{3}$ are functions of the boundary parameter $r$.

We can check that $e^{F_0(r)}$ plays a role of  the Bogoliubov
transformation
\begin{eqnarray*}
e^{-F_0(r)}h_{m}^{\ast 1}e^{F_0(r)} &=&h_{m}^{\ast 1}+\alpha
_{m}h_{-m}^{\ast 1}+\frac{\lbrack m\rbrack ^{2}}{m}\beta _{m}^{1}, \\
e^{-F_0(r)}h_{m}^{\ast 2}e^{F_0(r)} &=&h_{m}^{\ast 2}+\alpha
_{m}h_{-m}^{\ast 2}+\frac{\lbrack m\rbrack ^{2}}{m}\beta _{m}^{2},\\
e^{-F_0(r)}c_{m}e^{F_0(r)} &=&c_{m}+\alpha _{m}c_{-m}+\frac{%
\lbrack m\rbrack ^{2}}{m}\beta _{m}^{3}, \\
e^{-F_0(r)}h_{m}^{1}e^{F_0(r)} &=&h_{m}^{1}+\alpha
_{m}h_{-m}^{1}+\frac{\lbrack 2m\rbrack \lbrack m\rbrack }{m}\beta
_{m}^{1}-\beta _{m}^{2}\frac{\lbrack m\rbrack ^{2}}{m}.
\end{eqnarray*}
Keeping (\ref{Bst1}) in mind and following
\cite{Jim95,Miw97,Yan01},
we find that the coefficients $\alpha _{m},\beta
_{m}^{1},\beta _{m}^{2},\beta _{m}^{3}$ are
\bea
\alpha _{m} &=&-q^{4m},~~~~~\beta _{m}^{1}=0,\\
\beta _{m}^{2} &=&\frac{r^{m}}{\lbrack m\rbrack }q^{\frac{3}{2}m}+\theta
_{m}\,\frac{q^{^{\frac{3}{2}m}}-q^{\frac{5}{2}m}}{\lbrack m\rbrack }, \\
\beta _{m}^{3} &=&\theta _{m}\,\frac{q^{2m}}{\lbrack m\rbrack},
\eea
where the function $\theta_m$ is defined by
\bea
\theta _{m}=\lt\{
\begin{array}{ll}
    1 & {\rm if}~~m~~{\rm is~even} \\
    0 & {\rm if}~~m~~{\rm is~odd}
\end{array}
\rt..\no
\eea
Moreover following \cite{Miw97} one can check that $\eta_0|0>_B=0$,
namely, the boundary state $|0>_B$
$\in V(\u_0)$,  as required. In the derivation, the
following relation are useful
\begin{eqnarray*}
e^{h_{1}^{\ast +}(\xi ^{-1}q^{2};-\frac{1}{2})}|0;r>_{B}
&=&e^{h_{1}^{\ast -}(\xi q^{2};-\frac{1}{2})}|0;r>_{B},\\
e^{-h_{1}^{+}(\omega q^{2};-\frac{1}{2})}|0;r>_{B} &=&(1-\omega
^{-2})(1-r\omega ^{-1})e^{-h_{1}^{-}(\omega ^{-1}q^{2};-\frac{1}{2}%
)}|0;r>_{B}, \\
e^{c^{+}(\omega q^{2};0)}|0>_{B} &=&(1-\omega ^{-2})e^{c^{-}(\omega
^{-1}q^{2};0)}|0;r>_{B},\\
e^{-h_{2}^{\ast +}(\xi ^{-1}q^{2};-\frac{1}{2})}|0>_{B} &=&(1-\omega
r)^{-1}e^{-h_{2}^{\ast -}(\xi q^{2};-\frac{1}{2})}|0;r>_{B}.
\end{eqnarray*}
Similarly, the dual state ${}_B<r;0|\in V^*(\u_0)$ can be
constructed
\bea
{}_{B}
< r;0|&=&<0|e^{G_0(r)},\\
G_0(r)&=&-\frac{1}{2}\sum_{m=1}^{\infty }q^{-2m}\frac{m}{\lbrack
m\rbrack ^{2}}\left\{ h_{m}^{1}h_{m}^{1\ast }+h_{m}^{2}h_{m}^{2\ast
}+c_{m}c_{m}\right\}\no \\
&&+\sum_{m=1}^{\infty }\left\{ \delta _{m}^{1}h_{m}^{1}+\delta
_{m}^{2}h_{m}^{2}+\delta _{m}^{3}c_{m}\right\},
\eea
where
\begin{eqnarray*}
\delta _{m}^{1} &=&0, \\
\delta _{m}^{2} &=&-\frac{r^{-m}q^{-\frac{m}{2}}}{\lbrack m\rbrack }+\theta
_{m}\left( \frac{q^{-\frac{m}{2}}+q^{-\frac{3}{2}m}}{\lbrack m\rbrack }%
\right), \\
\delta _{m}^{3} &=&\theta _{m}\left( \frac{q^{-m}}{\lbrack m\rbrack
}\right).
\end{eqnarray*}

\subsection{The general boundary states}
Noting that the boundary K-matrix $K(z)$ have the
following properties
\bea
K(z)|_{z=r}=\lt(
\begin{array}{lll}
1&&\\
&0&\\
&&0
\end{array}
\rt),
\eea
we may define $|-1;r>_{B}=\Phi_{1}(r^{-1})|0;r>_B|_{r\longrightarrow
rq^{-2}}$. One can check that such $|-1;r>_B$ satisfies (\ref{Bst1})
with $\a=-1$. Recursively, we can construct the general
boundary state $|\a;r>_B$ from $|0;r>_B$ by the following
recursive relations,
\bea
|\a;rq^2>_B=\Phi_1(r^{-1})|\a+1;r>_B,~~
|\a;r>_B=q^{-2\rho_1}\Phi^*_1(r^{-1}q^{2})|\a-1;rq^2>_B.\label{Reu1}
\eea
We have used the second invertibility relation (\ref{DEF2}).
Similarly, we can obtain the dual boundary states ${}_B<r;\a|$
from ${}_B<r;0|$ by the recursive relations,
\bea
{}_B<r;\a|\Phi^*_1(r)={}_B<rq^2;\a-1|,~~
{}_B<r;\a|={}_B<rq^2;\a-1|\Phi_1(rq^{-2})q^{-2\rho_1}.\label{Reu2}
\eea

\sect{Correlation functions}

The aim of this section is to calculate the one-point
functions $<E^{(1)}_{i,j}>_{\a}$:
\bea
<E^{(1)}_{i,j}>_{\a}=\frac{{}_B<r;\a|E^{(1)}_{i,j}|\a;r>_B}{{}_B<r;\a|\a;r>_B}.\no
\eea
The
generalization to the calculation of multi-point functions is straightforward.
Thanks to the recursive relations (\ref{Reu1}) and (\ref{Reu2}),
it is sufficient to calculate $<E^{(1)}_{i,j}>_0$.
Thus in the following we restrict ourselves to the
calculation  of $<E^{(1)}_{i,j}>_0$.

Define
\bea
\oint dz~f(z)=f_{-1},~~{\rm for~ formal~series~function~}f(z)=
\sum_{n\in Z}f_nz^n\no.
\eea
By the bosonic realization of Drinfeld currents of
$\uqs2h$, (\ref{VOX1})-(\ref{VOX2}) and the normal ordering relations in
the appendix A, we obtain the integral expression
of the vertex operators \cite{Yan99}
\begin{eqnarray*}
\phi_3(z)&=&:e^{-h^{*}_2(q^2z;-\frac{1}{2})+c(q^2z;0)}:
  e^{-i\pi  a^{2}_{0}},\\
\phi_2(z)&=&\{\frac{e^{-c(wq;0)}}{wq(1-\frac{qz}{w})}
 +\frac{e^{-c(wq^{-1};0)}}{zq^2(1-\frac{w}{zq^3})}\}
 e^{-h^{*}_2(q^2z;-\frac{1}{2})-h_2(w;-\frac{1}{2})+c(q^2z;0)}
 e^{i\pi h^1_0}: ,\\
\phi_1(z)&=&\frac{q^2-1}{w(1-\frac{w_1q}{w})(1-\frac{wq}{w_1})}\\
& & \times :\{\frac{e^{-c(wq;0)}}{wq(1-\frac{qz}{w})}
  +\frac{e^{-c(wq^{-1};0)}}{zq^2(1-\frac{w}{zq^3})}\}
 e^{-h^{*}_2(q^2z;-\frac{1}{2})-h_2(w;-\frac{1}{2})
 -h_1(w_1;-\frac{1}{2})+c(q^2z;0)}
 e^{-i\pi a^2_0}: ,\\
\phi^{*}_1(z)&=&:e^{h^{*}_1(qz;-\frac{1}{2})}:e^{i\pi a^{2}_{0}},\\
\phi^{*}_2(z)&=&\oint dw\frac{1-q^{-2}}{z(1-\frac{zq^2}{w})
 (1-\frac{w}{z})}:e^{h^{*}_1(qz;-\frac{1}{2})-h_1(w;-\frac{1}{2})}
 e^{-i\pi h^{1}_0}: ,\\
\phi^{*}_3(z)&=&\oint dw_1\oint dw \frac{1-q^{-2}}{z(1-\frac{zq^2}{w})
 (1-\frac{w}{z})}\\
& & \times :\frac{e^{-c(w_1q;0)}-e^{-c(w_1q^{-1};0)}}
 {ww_1(1-\frac{wq}{w_1})(1-\frac{w_1q}{w})}
 e^{h^{*}_1(qz;-\frac{1}{2})-h_1(w;-\frac{1}{2})-h_2(w_1;-\frac{1}{2})}
 e^{i\pi a^{2}_0}:
\end{eqnarray*}

Since $\eta_0|0,r>=0$, one may set
\bea
P_{i,j}(z_1,z_2)=\frac{{}_B<r;0|\P^*_{i}(z_1)\P_j(z_2)|0;r>_B}{{}<r;0|0;r>_B}
\equiv
\frac{{}_B<r;0|\phi^*_{i}(z_1)\phi_j(z_2)|0;r>_B}{{}<r;0|0;r>_B},
\eea
then $<E^{(1)}_{i,j}>_0=-(-1)^{[j]}P_{i,j}(1,1)$.

The bosonization formulae (\ref{VOX1})-(\ref{VOX2})
of the vertex operators immediately imply
\bea
&&P_{i,j}(z_1,z_2)=\d_{ij}F_i(z_1,z_2)
\stackrel{def}{=}
\d_{ij}\frac{{}_B<r;0|\phi^*_{i}(z_1)\phi_i(z_2)|0;r>_B}{{}_B<r;0|0;r>_B}.\no
\eea
Using the technique in \cite{Jim95,Yan01} (see equation (\ref{ID})),
after tedious calculation, we get
\begin{eqnarray}
_{B}\langle r;0|0;r\rangle _{B} &=&\prod_{n=1}^{\infty }\frac{1}{1-\alpha
_{n}\gamma _{n}}\prod_{n=1}^{\infty }\frac{1}{\left( \alpha _{n}\gamma
_{n}-1\right) ^{\frac{1}{2}}} \no\\
&&\exp \left[ \frac{1}{2}\sum \frac{\lbrack n\rbrack ^{2}}{n}\frac{1}{%
1-\alpha _{n}\gamma _{n}}(\gamma _{n}\left( \beta _{n}^{3}\right)
^{2}+2\beta _{n}^{3}\delta _{n}^{3}+\alpha _{n}\left( \delta _{n}^{3}\right)
^{2}\right],
\end{eqnarray}

\begin{eqnarray}
F_{1}(z_1,z_2) &=&\frac{1}{_{B}\langle r;0|0;r\rangle _{B}}
\oint d\omega _{1}\oint d\omega \frac{(q^{2}-1)g_1}{q\omega ^{2}(1-%
\frac{\omega _{1}q}{\omega })(1-\frac{\omega q}{\omega _{1}})(1-\frac{z_{2}q%
}{\omega })} \no\\
&&\times~\prod_{n=1}^{\infty }(-(\alpha _{n}
\gamma _{n}-1)^{-1})\prod_{n=1}^{\infty
}(\alpha _{n}\gamma _{n}-1)^{-\frac{1}{2}} \no\\
&&\times~\exp (\sum \frac{\lbrack n\rbrack ^{2}}{n}\frac{1}{(\alpha _{n}\gamma
_{n}-1)}\{(B_1-C_1)^{2}\frac{\lbrack 2n\rbrack }{\lbrack n\rbrack }\frac{\gamma
_{n}}{\alpha _{n}\gamma _{n}-1}+\gamma _{n}(B_1-C_1)\beta _{n}^{2}\no \\
&&~~-\gamma _{n}(B_1-C_1)A_1+(B_1-C_1)\delta _{n}^{2}\}) \no\\
&&\times~\exp (\sum \frac{\lbrack n\rbrack ^{2}}{n}\frac{1}{1-\alpha _{n}
\gamma _{n}%
}\{\frac{1}{2}(\beta _{n}^{3})^{2}D\gamma _{n}+\frac{1}{2}\beta
_{n}^{3}D_1^{2}\gamma _{n}+\beta _{n}^{3}D_1\gamma _{n}+\beta _{n}^{3}\delta
_{n}^{3}\no \\
&&~~~+D_1\delta _{n}^{3}+\frac{1}{2}\alpha _{n}\left( \delta _{n}^{3}\right)
^{2}\}) \no\\
&&+\oint d\omega _{1}\oint d\omega \frac{(q^{2}-1)g_1^{\prime }}{q^{2}\omega
z_{2}(1-\frac{\omega _{1}q}{\omega })(1-\frac{\omega q}{\omega _{1}})(1-%
\frac{\omega }{z_{2}q^{3}})}\no \\
&&\times~\prod_{n=1}^{\infty }
(-(\alpha _{n}\gamma _{n}-1)^{-1})\prod_{n=1}^{\infty
}(\alpha _{n}\gamma _{n}-1)^{-\frac{1}{2}} \no\\
&&\times~\exp (\sum \frac{\lbrack n\rbrack ^{2}}{n}\frac{1}{(\alpha _{n}\gamma
_{n}-1)}\{(B_1-C_1)^{2}\frac{\lbrack 2n\rbrack }{\lbrack n\rbrack }\frac{\gamma
_{n}}{\alpha _{n}\gamma _{n}-1}+\gamma _{n}(B_1-C_1)\beta _{n}^{2}\no \\
&&-\gamma _{n}(B_1-C_1)A_1+(B_1-C_1)\delta _{n}^{2}\})\no \\
&&\times\exp (\sum \frac{\lbrack n\rbrack ^{2}}{n}
\frac{1}{1-\alpha _{n}\gamma _{n}%
}\{\frac{1}{2}(\beta _{n}^{3})^{2}D_1^{\prime }\gamma _{n}+\frac{1}{2}\beta
_{n}^{3}(D_1^{\prime })^{2}\gamma _{n}+\beta _{n}^{3}D_1^{\prime }\gamma
_{n}+\beta _{n}^{3}\delta _{n}^{3}\no \\
&&+D_1^{\prime }\delta _{n}^{3}+\frac{1}{2}\alpha _{n}\left( \delta
_{n}^{3}\right) ^{2}\}),
\end{eqnarray}
where
\begin{eqnarray*}
g_1 &=&\exp \left( -\sum \frac{q^{3n}z_{2}^{-n}\omega ^{-n}}{n}\right) \exp
\left( \sum \frac{q^{n}z_{2}^{-n}\omega ^{-n}}{n}\right) \exp \left( \sum
\frac{q^{n}z_{2}^{n}\omega ^{-n}}{n}\right) \\
&&\exp \left( -\sum \frac{q^{-n}z_{2}^{-n}\omega ^{n}}{n}\right) \exp \left(
\sum \frac{r^{n}z_{2}^{-n}}{n}\right) \exp \left( \sum \frac{q^{5n}\omega
^{-n}\omega _{1}^{-n}}{n}\right) \\
&&\exp \left( \sum \frac{q^{4n}z_{1}^{-n}\omega _{1}^{-n}}{n}\right) \exp
\left( \sum \frac{q^{n}\omega ^{-n}\omega _{1}^{n}}{n}\right) \exp \left(
-\sum \frac{q^{4n}\omega _{1}^{-2n}}{n}\right) \\
&&\exp \left( -\sum \frac{r^{n}q^{2n}\omega _{1}^{-n}}{n}\right) \exp \left(
-\sum \frac{q^{2n}z_{2}^{-n}z_{1}^{-n}}{n}\right) \exp \left( -\sum \frac{%
q^{2n}z_{2}^{n}z_{1}^{-n}}{n}\right) \\
&&\exp \left( \sum \frac{z_{1}^{-n}\omega
_{1}^{n}}{n}\right),
\end{eqnarray*}
\begin{eqnarray*}
g_1^{^{\prime }} &=&\exp \left( \sum \frac{q^{n}z_{2}^{n}\omega ^{-n}}{n}%
\right) \exp \left( -\sum \frac{q^{-n}z_{2}^{-n}\omega ^{n}}{n}\right) \exp
\left( \sum \frac{r^{n}z_{2}^{-n}}{n}\right) \\
&&\times\exp \left( \sum \frac{q^{5n}\omega ^{-n}
\omega _{1}^{-n}}{n}\right) \exp
\left( \sum \frac{q^{4n}z_{1}^{-n}\omega _{1}^{-n}}{n}\right) \exp \left(
\sum \frac{q^{n}\omega ^{-n}\omega _{1}^{n}}{n}\right) \\
&&\times\exp \left( -\sum \frac{q^{4n}\omega _{1}^{-2n}}{n}\right) \exp \left(
-\sum \frac{r^{n}q^{2n}\omega _{1}^{-n}}{n}\right) \exp \left( -\sum \frac{%
q^{2n}z_{2}^{-n}z_{1}^{-n}}{n}\right) \\
&&\times\exp \left( -\sum \frac{q^{2n}z_{2}^{n}z_{1}^{-n}}{n}\right) 
\exp \left(
\sum \frac{z_{1}^{-n}\omega _{1}^{n}}{n}\right),
\end{eqnarray*}
and
\begin{eqnarray*}
A_1 &=&\sum \frac{q^{\frac{3}{2}n}z_{1}^{n}}{\lbrack n\rbrack }-\alpha
_{n}\sum \frac{q^{-\frac{1}{2}n}z_{1}^{-n}}{\lbrack n\rbrack }+\sum \frac{q^{%
\frac{1}{2}n}\omega ^{n}}{\lbrack n\rbrack }-\alpha _{n}\sum \frac{q^{\frac{1%
}{2}n}\omega ^{-n}}{\lbrack n\rbrack }, \\
B_1 &=&\sum \frac{q^{\frac{5}{2}n}z_{2}^{n}}{\lbrack n\rbrack }-\alpha
_{n}\sum \frac{q^{-\frac{3}{2}n}z_{2}^{-n}}{\lbrack n\rbrack }, \\
D_1 &=&\sum \frac{q^{2n}z_{2}^{n}}{\lbrack n\rbrack }-\alpha _{n}\sum \frac{%
q^{-2n}z_{2}^{-n}}{\lbrack n\rbrack }-\sum \frac{q^{n}\omega ^{n}}{\lbrack
n\rbrack }+\alpha _{n}\sum \frac{q^{-n}\omega ^{-n}}{\lbrack n\rbrack }, \\
C_1 &=&\sum \frac{q^{\frac{1}{2}n}\omega _{1}^{n}}{\lbrack n\rbrack }-\alpha
_{n}\sum \frac{q^{\frac{1}{2}n}\omega _{1}^{-n}}{\lbrack n\rbrack }, \\
D_1^{^{\prime }} &=&\sum \frac{q^{2n}z_{2}^{n}}{\lbrack n\rbrack }-\alpha
_{n}\sum \frac{q^{-2n}z_{2}^{-n}}{\lbrack n\rbrack }-\sum \frac{q^{-n}\omega
^{n}}{\lbrack n\rbrack }+\alpha _{n}\sum \frac{q^{n}\omega ^{-n}}{\lbrack
n\rbrack },
\end{eqnarray*}
\begin{eqnarray}
F_{2}(z_1,z_2) &=&\frac{1}{_{B}\langle r;0|0;r\rangle _{B}}
\oint d\omega \oint d\omega _{1}\frac{\left( 1-q^{-2}\right) g_2}{%
z_{1}\left( 1-\frac{z_{1}q^{2}}{\omega }\right) \left( 1-\frac{\omega }{z_{1}%
}\right) \omega _{1}q\left( 1-\frac{z_{2}q}{\omega _{1}}\right) }\no \\
&&\times \prod_{n=1}^{\infty }(-(\alpha _{n}\gamma _{n}-1)^{-1})
\prod_{n=1}^{\infty
}(\alpha _{n}\gamma _{n}-1)^{-\frac{1}{2}}\no \\
&&\times \exp \{\sum \frac{\lbrack n\rbrack ^{2}}{n}\frac{1}{\alpha _{n}\gamma
_{n}-1}\lbrack (B_2-C_2)^{2}\frac{\lbrack 2n\rbrack }{\lbrack n\rbrack }\frac{%
\gamma _{n}}{\alpha _{n}\gamma _{n}-1}+\gamma _{n}(B_2-C_2)\beta _{n}^{2}\no \\
&&~~~-\gamma _{n}(B_2-C_2)A_2+(B_2-C_2)\delta _{n}^{2}\rbrack \}\no \\
&&\times\exp \{\sum \frac{\lbrack n\rbrack ^{2}}{n}\frac{1}{1-\alpha _{n}\gamma
_{n}}\lbrack \frac{1}{2}\left( \beta _{n}^{3}\right) ^{2}D_2\gamma _{n}+\frac{1%
}{2}\beta _{n}^{3}D_2^{2}\gamma _{n}+\beta _{n}^{3}D_2\gamma _{n}+\beta
_{n}^{3}\delta _{n}^{3}\no \\
&&~~~+D_2\delta _{n}^{3}+\frac{1}{2}\alpha _{n}\left( \delta _{n}^{3}\right)
^{2}\rbrack \}\no \\
&&+\oint d\omega \oint d\omega _{1}\frac{\left( 1-q^{-2}\right) g_2^{^{\prime
}}}{z_{1}\left( 1-\frac{z_{1}q^{2}}{\omega }\right) \left( 1-\frac{\omega }{%
z_{1}}\right) z_{2}q^{2}\left( 1-\frac{\omega _{1}}{z_{2}q^{3}}\right) }\no \\
&&\times\exp \{\sum \frac{\lbrack n\rbrack ^{2}}{n}\frac{1}{\alpha _{n}\gamma
_{n}-1}\lbrack (B_2-C_2)^{2}\frac{\lbrack 2n\rbrack }{\lbrack n\rbrack }\frac{%
\gamma _{n}}{\alpha _{n}\gamma _{n}-1}+\gamma _{n}(B_2-C_2)\beta _{n}^{2}\no \\
&&~~~-\gamma _{n}(B_2-C_2)A_2+(B_2-C_2)\delta _{n}^{2}\rbrack \}\no \\
&&\times\prod_{n=1}^{\infty }(-(\alpha _{n}\gamma
_{n}-1)^{-1})\prod_{n=1}^{\infty}(\alpha _{n}
\gamma _{n}-1)^{-\frac{1}{2}} \no\\
&&\times \exp \{\sum \frac{\lbrack n\rbrack ^{2}}{n}
\frac{1}{1-\alpha _{n}\gamma
_{n}}\lbrack \frac{1}{2}\left( \beta _{n}^{3}\right) ^{2}D_2^{^{\prime
}}\gamma _{n}+\frac{1}{2}\beta _{n}^{3}D_2^{^{\prime }2}\gamma _{n}+\beta
_{n}^{3}D_2^{^{\prime }}\gamma _{n}+\beta _{n}^{3}\delta _{n}^{3}\no \\
&&+~~~D_2^{^{\prime }}\delta _{n}^{3}+\frac{1}{2}\alpha _{n}\left( \delta
_{n}^{3}\right) ^{2}\rbrack \},
\end{eqnarray}
where
\begin{eqnarray*}
g_2 &=&\exp \left( \sum \frac{\omega ^{n}z_{1}^{-n}}{n}\right) \exp \left(
-\sum \frac{q^{2n}z_{1}^{-n}z_{2}^{n}}{n}\right) \exp \left( \sum \frac{%
q^{n}\omega ^{-n}\omega _{1}^{n}}{n}\right)  \\
&&\times\exp \left( -\sum \frac{q^{-n}\omega _{1}^{n}z_{2}^{-n}}{n}\right) \exp
\left( \sum \frac{q^{n}\omega _{1}^{-n}z_{2}^{n}}{n}\right) \exp \left( \sum
\frac{r^{n}z_{2}^{-n}}{n}\right)  \\
&&\times\exp \left( -\sum 
\frac{\omega ^{-2n}q^{4n}}{n}\right) \exp \left( -\sum
\frac{\omega ^{-n}q^{2n}r^{n}}{n}\right) \exp \left( \sum \frac{q^{4n}\omega
^{-n}z_{1}^{-n}}{n}\right)  \\
&&\times\exp \left( \sum \frac{q^{5n}
\omega ^{-n}\omega _{1}^{-n}}{n}\right) \exp
\left( -\sum \frac{q^{3n}\omega _{1}^{-n}z_{2}^{-n}}{n}\right) \exp \left(
\sum \frac{q^{n}\omega _{1}^{-n}z_{2}^{-n}}{n}\right)  \\
&&\times\exp \left( -\sum
\frac{q^{2n}z_{2}^{-n}z_{1}^{-n}}{n}\right),
\end{eqnarray*}
\begin{eqnarray*}
g_2^{^{\prime }} &=&\exp \left( \sum \frac{\omega ^{n}z_{1}^{-n}}{n}\right)
\exp \left( -\sum \frac{q^{2n}z_{2}^{n}z_{1}^{-n}}{n}\right) \exp \left(
\sum \frac{q^{n}\omega ^{-n}\omega _{1}^{n}}{n}\right)	\\
&&\times\exp \left( -\sum \frac{q^{-n}\omega _{1}^{n}z_{2}^{-n}}{n}\right) \exp
\left( \sum \frac{q^{3n}\omega _{1}^{-n}z_{2}^{n}}{n}\right) \exp \left(
\sum \frac{r^{n}z_{2}^{-n}}{n}\right)  \\
&&\times\exp \left( -\sum 
\frac{\omega ^{-2n}q^{4n}}{n}\right) \exp \left( -\sum
\frac{\omega ^{-n}q^{2n}r^{n}}{n}\right) \exp \left( \sum \frac{q^{4n}\omega
^{-n}z_{1}^{-n}}{n}\right)  \\
&&\times\exp \left( \sum \frac{q^{5n}
\omega ^{-n}\omega _{1}^{-n}}{n}\right) \exp
\left( -\sum \frac{q^{2n}z_{2}^{-n}z_{1}^{-n}}{n}\right),
\end{eqnarray*}
and
\begin{eqnarray*}
A_2 &=&\sum \frac{q^{\frac{3}{2}n}z_{1}^{n}}{\lbrack n\rbrack }-\alpha
_{n}\sum \frac{q^{-\frac{1}{2}n}z_{1}^{-n}}{\lbrack n\rbrack }+\sum \frac{q^{%
\frac{1}{2}n}\omega _{1}^{n}}{\lbrack n\rbrack }-\alpha _{n}\sum \frac{q^{%
\frac{1}{2}n}\omega _{1}^{-n}}{\lbrack n\rbrack }, \\
B_2 &=&\sum \frac{q^{\frac{5}{2}n}z_{2}^{n}}{\lbrack n\rbrack }-\alpha
_{n}\sum \frac{q^{-\frac{3}{2}n}z_{2}^{-n}}{\lbrack n\rbrack }, \\
D_2 &=&\sum \frac{q^{2n}z_{2}^{n}}{\lbrack n\rbrack }-\alpha _{n}\sum \frac{%
q^{-2n}z_{2}^{-n}}{\lbrack n\rbrack }-\sum \frac{q^{n}\omega _{1}^{n}}{%
\lbrack n\rbrack }+\alpha _{n}\sum \frac{q^{-n}\omega _{1}^{-n}}{\lbrack
n\rbrack }, \\
C_2 &=&\sum \frac{q^{\frac{1}{2}n}\omega _{1}^{n}}{\lbrack n\rbrack }-\alpha
_{n}\sum \frac{q^{\frac{1}{2}n}\omega _{1}^{-n}}{\lbrack n\rbrack }, \\
D_2^{^{\prime }} &=&\sum \frac{q^{2n}z_{2}^{n}}{\lbrack n\rbrack }-\alpha
_{n}\sum \frac{q^{-2n}z_{2}^{-n}}{\lbrack n\rbrack }-\sum \frac{q^{-n}\omega
_{1}^{n}}{\lbrack n\rbrack }+\alpha _{n}\sum \frac{q^{n}\omega _{1}^{-n}}{%
\lbrack n\rbrack },
\end{eqnarray*}
\begin{eqnarray}
F_{3}(z_1, z_2) &=&\frac{-1}{_{B}\langle r;0|0;r\rangle _{B}}
\oint d\omega \oint d\omega _{1}\frac{\left( 1-q^{-2}\right) g_3}{%
z_{1}\left( 1-\frac{z_{1}q^{2}}{\omega }\right) \left( 1-\frac{\omega }{z_{1}%
}\right) \omega _{1}\omega \left( 1-\frac{\omega q}{\omega _{1}}\right)
\left( 1-\frac{\omega _{1}q}{\omega }\right) }\no \\
&&\times\prod_{n=1}^{\infty }(-(\alpha _{n}\gamma
_{n}-1)^{-1})\prod_{n=1}^{\infty}(\alpha _{n}
\gamma _{n}-1)^{-\frac{1}{2}} \no\\
&&\times\exp \{\sum \frac{\lbrack n\rbrack ^{2}}{n}\frac{1}{\alpha _{n}\gamma
_{n}-1}\lbrack (B_3-C_3)^{2}\frac{\lbrack 2n\rbrack }{\lbrack n\rbrack }\frac{%
\gamma _{n}}{\alpha _{n}\gamma _{n}-1}+\gamma _{n}(B_3-C_3)\beta _{n}^{2}\no \\
&&~~~-\gamma _{n}(B_3-C_3)A_3+(B_3-C_3)\delta _{n}^{2}\rbrack \}\no \\
&&\times \exp \{\sum \frac{\lbrack n\rbrack ^{2}}{n}
\frac{1}{1-\alpha _{n}\gamma
_{n}}\lbrack \frac{1}{2}
\left( \beta _{n}^{3}\right) ^{2}D_3\gamma _{n}+\frac{1%
}{2}\beta _{n}^{3}D_3^{2}\gamma _{n}+\beta _{n}^{3}D_3\gamma _{n}+\beta
_{n}^{3}\delta _{n}^{3}\no \\
&&+D_3\delta _{n}^{3}+\frac{1}{2}\alpha _{n}\left( \delta _{n}^{3}\right)
^{2}\rbrack \}\no \\
&&+\oint d\omega \oint d\omega _{1}\frac{\left( 1-q^{-2}\right) g_3^{^{\prime
}}}{z_{1}\left( 1-\frac{z_{1}q^{2}}{\omega }\right) \left( 1-\frac{\omega }{%
z_{1}}\right) \omega \omega _{1}\left( 1-\frac{\omega q}{\omega _{1}}\right)
\left( 1-\frac{\omega _{1}q}{\omega }\right) }\no \\
&&\times \prod_{n=1}^{\infty }(-(\alpha _{n}\gamma
_{n}-1)^{-1})\prod_{n=1}^{\infty}(\alpha _{n}
\gamma _{n}-1)^{-\frac{1}{2}}\no \\
&&\times\exp \{\sum \frac{\lbrack n\rbrack ^{2}}{n}\frac{1}{\alpha _{n}\gamma
_{n}-1}\lbrack (B_3-C_3)^{2}\frac{\lbrack 2n\rbrack }{\lbrack n\rbrack }\frac{%
\gamma _{n}}{\alpha _{n}\gamma _{n}-1}
+\gamma _{n}(B_3-C_3)\beta _{n}^{2}\no \\
&&~~~-\gamma _{n}(B_3-C_3)A_3+(B_3-C_3)\delta _{n}^{2}\rbrack \}\no \\
&&\times \exp \{\sum \frac{\lbrack n\rbrack ^{2}}{n}
\frac{1}{1-\alpha _{n}\gamma
_{n}}\lbrack \frac{1}{2}\left( \beta _{n}^{3}\right) ^{2}D_3^{^{\prime
}}\gamma _{n}+\frac{1}{2}\beta _{n}^{3}D^{^{\prime }2}_3\gamma _{n}+\beta
_{n}^{3}D_3^{^{\prime }}\gamma _{n}+\beta _{n}^{3}\delta _{n}^{3}\no \\
&&~~~+D_3^{^{\prime }}\delta _{n}^{3}+\frac{1}{2}\alpha _{n}\left( \delta
_{n}^{3}\right) ^{2}\rbrack \},
\end{eqnarray}
where
\begin{eqnarray*}
g_3 &=&\exp \left( -\sum \frac{\omega ^{n}z_{1}^{-n}}{n}\right) \left( -\sum
\frac{q^{2n}z_{1}^{-n}z_{2}^{n}}{n}\right) \exp \left( \sum \frac{%
q^{n}\omega ^{-n}\omega _{1}^{n}}{n}\right)  \\
&&\times\exp \left( -\sum \frac{q^{3n}\omega
_{1}^{-n}z_{2}^{n}}{n}\right) \exp
\left( \sum \frac{q^{n}\omega _{1}^{-n}z_{2}^{n}}{n}\right) \exp \left( \sum
\frac{r^{n}z_{2}^{-n}}{n}\right)  \\
&&\times\exp \left( -\sum 
\frac{\omega ^{-2n}q^{4n}}{n}\right) \exp \left( -\sum
\frac{\omega ^{-n}q^{2n}r^{n}}{n}\right) \exp \left( \sum \frac{q^{4n}\omega
^{-n}z_{1}^{-n}}{n}\right)  \\
&&\times\exp \left( \sum 
\frac{q^{5n}\omega ^{-n}\omega _{1}^{-n}}{n}\right) \exp
\left( -\sum \frac{q^{3n}\omega _{1}^{-n}z_{2}^{-n}}{n}\right) \exp \left(
\sum \frac{q^{n}\omega _{1}^{-n}z_{2}^{-n}}{n}\right)  \\
&&\times\exp \left( -\sum \frac{q^{2n}z_{2}^{-n}z_{1}^{-n}}{n}\right),	\\
g_3^{^{\prime }} &=&\exp \left( -\sum \frac{\omega ^{n}z_{1}^{-n}}{n}\right)
\exp \left( \sum \frac{q^{n}\omega ^{-n}\omega _{1}^{n}}{n}\right) \left(
-\sum \frac{q^{2n}z_{1}^{-n}z_{2}^{n}}{n}\right)  \\
&&\times\exp \left( \sum 
\frac{r^{n}z_{2}^{-n}}{n}\right) \exp \left( -\sum \frac{%
\omega ^{-2n}q^{4n}}{n}\right) \exp \left( -\sum \frac{\omega
^{-n}q^{2n}r^{n}}{n}\right)  \\
&&\times\exp \left( -\sum 
\frac{q^{2n}z_{2}^{-n}z_{1}^{-n}}{n}\right) \exp \left(
\sum \frac{q^{5n}\omega ^{-n}\omega _{1}^{-n}}{n}\right) \exp \left( \sum
\frac{q^{4n}\omega ^{-n}z_{1}^{-n}}{n}\right),
\end{eqnarray*}
and
\begin{eqnarray*}
A_3 &=&\sum \frac{q^{\frac{3}{2}n}z_{1}^{n}}{\lbrack n\rbrack }-\alpha
_{n}\sum \frac{q^{-\frac{1}{2}n}z_{1}^{-n}}{\lbrack n\rbrack }+\sum \frac{q^{%
\frac{1}{2}n}\omega _{1}^{n}}{\lbrack n\rbrack }-\alpha _{n}\sum \frac{q^{%
\frac{1}{2}n}\omega _{1}^{-n}}{\lbrack n\rbrack }, \\
B_3 &=&\sum \frac{q^{\frac{5}{2}n}z_{2}^{n}}{\lbrack n\rbrack }-\alpha
_{n}\sum \frac{q^{-\frac{3}{2}n}z_{2}^{-n}}{\lbrack n\rbrack }, \\
D_3 &=&\sum \frac{q^{2n}z_{2}^{n}}{\lbrack n\rbrack }-\alpha _{n}\sum \frac{%
q^{-2n}z_{2}^{-n}}{\lbrack n\rbrack }-\sum \frac{q^{n}\omega _{1}^{n}}{%
\lbrack n\rbrack }+\alpha _{n}\sum \frac{q^{-n}\omega _{1}^{-n}}{\lbrack
n\rbrack }, \\
C_3 &=&\sum \frac{q^{\frac{1}{2}n}\omega _{1}^{n}}{\lbrack n\rbrack }-\alpha
_{n}\sum \frac{q^{\frac{1}{2}n}\omega _{1}^{-n}}{\lbrack n\rbrack }, \\
D_3^{^{\prime }} &=&\sum \frac{q^{2n}z_{2}^{n}}{\lbrack n\rbrack }-\alpha
_{n}\sum \frac{q^{-2n}z_{2}^{-n}}{\lbrack n\rbrack }-\sum \frac{q^{-n}\omega
_{1}^{n}}{\lbrack n\rbrack }+\alpha _{n}\sum \frac{q^{n}\omega _{1}^{-n}}{%
\lbrack n\rbrack }.
\end{eqnarray*}

We now derive the difference equations satisfied by the one-point
functions. By (\ref{TF}) and
(\ref{DEF1})-(\ref{DEF2}), one obtains
\bea
\P^*_i(z^{-1})|\a;r>_B&=&
 \sum_{j}\bar{K}^j_i(zq)\P^*_j(zq^2)|\a;r>_B,\label{Bst3}\\
{}_B<r;\a|\P_i(z)(-1)^{[i]}&=&
 \sum_{j}{}_B<r;\a|\P_j(z^{-1}q^{-2})\bar{K}^j_i(zq)
(-1)^{[j]}.\label{Bst4}
\eea
By (\ref{ZF3}),
one derives the exchange relations
\bea
\Phi^{*}_i(z_1)\Phi_j(z_2)=\sum_{kl}
\stackrel{\sim}{R}(\frac{z_1}{z_2})_{ij}^{kl}
\Phi_l(z_2)\Phi^{*}_k(z_1)(-1)^{[k][l]},\label{ZF4}
\eea
where $\stackrel{\sim}{R}(z)=R^{-1,st_1,-1}(z)$.

Using (\ref{Bst1})-(\ref{Bst2}), (\ref{Bst3})-(\ref{ZF4}),
(\ref{ZF3}) and
(\ref{VOX1})-(\ref{VOX2}), we get the difference equations
\bea
F_i(z_1q^{-2},z_2)&=&\sum_{j,k,l,m,n}(-1)^{
[k][l]+[i]+[j]+[n]}K^i_j(z_1q^{-2})
\stackrel{\sim}{R}(z_1^{-1}z_2^{-1}q^2)^{lk}_{ji}\no\\
&&~~~~~~~~~~~~~~~~~~~\times \bar{K}^{m}_{l}(z_1q^{-1})
\bar{R}(\frac{z_1}{z_2})^{n~ n}_{m~k}~F_{n}(z_1,z_2),\\
F_i(z_1,z_2q^2)&=&\sum_{j,k,l,m,n}(-1)^{[k][l]+[l]+[m]+[n]}
K_i^j(z_2^{-1}q^{-2})\stackrel{\sim}{R}(z_1z_2q^2)^{ij}_{kl}\no\\
&&~~~~~~~~~~~~~~~~~~~\times \bar{K}^{m}_{l}(z_2^{-1}q^{-1})
\bar{R}(\frac{z_1}{z_2})^{n~n}_{k~m}~F_{n}(z_1,z_2).
\eea

\vskip 1cm
\noindent{\Large \bf Acknowledgment}
\vskip 0.8cm

\noindent
We would like to thank G. von Gehlen for his interest in this
problem. This work
has been partly supported by
the grant of National Natural Science Foundation of China.
W.-L. Yang is supported by the
Alexander von Humboldt Foundation. Y.-Z. Zhang is supported by
Australian Research Council.

\appendix

\sect{Appendix A}
\subsection{Bosonization of $\uqs2h$}

In this section, we briefly review the bosonization of
$\uqs2h$ at level-one and the
corresponding vertex operators \cite{Kim97,Yan99}.

The Cartan matrix of $\uqs2h$ is
\begin{eqnarray*}
(a_{ij})=\left(
\begin{array}{ccc}
0&-1&1\\
-1&2&-1\\
1&-1&0
\end{array}\right)
\end{eqnarray*}
where $i,j=0,1,2$. 

In terms of the Drinfeld generators: $\{d,\ \ X^{\pm,i}_{m}$,
$h^{i}_n$, $(K^i)^{\pm 1},\gamma^{\pm 1/2} | i=1,2,$ $m \in
{\bf Z}, n \in {\bf Z}_{\ne 0}\}$, the defining	relations of $\uqs2h$ read
\begin{eqnarray*}
& &\gamma\ \  {\rm is\ \  central },~~
[K^i,h^j_m]=0,~~[d,K^i]=0,~~ [d,h^{j}_m]=mh^j_m,\\
& &[h^i_m,h^j_n] =\delta_{m+n,0}
{[a_{ij}m](\gamma^m-\gamma^{-m}) \over m(q-q^{-1})}, \\
& &K^iX^{\pm,j}_m =q^{\pm a_{ij}}X^{\pm,j}_m K^i,~~
[d,X^{\pm,j}_m]=mX^{\pm,j}_m ,\\
& &[h^i_m,X^{\pm,j}_n]=\pm {[a_{ij}m] \over m}
\gamma^{\pm |m|/2}X^{\pm,j}_{n+m},\\
& &[X^{+,i}_m,X^{-,j}_n]=\frac{\delta_{i,j}}{ q-q^{-1}}
(\gamma^{(m-n)/2}\psi^{+,j}_{m+n} -\gamma^{-(m-n)/2 }
\psi^{-,j}_{m+n}),\\
& &[X^{\pm ,2}_m,X^{\pm ,2}_n]=0, \\
& &[X^{\pm,i}_{m+1},X^{\pm,j}_{n}]_{q^{\pm a_{ij}}}
+[X^{\pm,j}_{n+1},X^{\pm,i}_{m}]_{q^{\pm a_{ij}}}=0,
\ \ {\rm for} \ \ a_{ij}\ne 0,
\end{eqnarray*}
where $[m]=\frac{q^m-q^{-m}}{q-q^{-1}}$,
$[X,Y]_\xi=XY-(-1)^{[X][Y]}\xi YX$ and 
$[X,Y]_1 \equiv [X,Y]$; 
the ${\bf Z}_2$-grading of Drinfeld generators
are : $[X^{\pm,2}_m]=1$ for $m\in {\bf Z}$ and zero otherwise.

Introduce the bosonic q-oscillators \cite{Kim97}
$\{a^1_n,a^2_n,b_n,c_n$, $Q_{a^1},Q_{a^2}, Q_{b},Q_{c}$
$|n \in {\bf Z} \}$, which satisfy the commutation relations
\begin{eqnarray*}
& &[a^i_m,a^j_n]=\delta_{i,j}\delta_{m+n,0}{[m]^2 \over
m},~~
[a^i_0,Q_{a^j}] = \delta_{i,j},\\
& &[b_m,b_n]=-\delta_{m+n,0}{[m]^2 \over m},~~
[b_0,Q_{b}] = -1,  \\
& &[c_m,c_n]=\delta_{m+n,0}{[m]^2 \over m},~~
[c_0,Q_{c}] = 1.
\end{eqnarray*}
Define the generating functions for the Drinfeld basis by
$X^{\pm}_i(z)=\sum_{m\in {\bf Z}}X^{\pm,i}_mz^{-m-1} $,
and introduce $h^i_0$ by setting $K^i=q^{h^i_0}$.
Define
$Q_{h^1}=Q_{a^1}-Q_{a^{2}}$,
$Q_{h^{2}}=Q_{a^{2}}+Q_{b}$ and $h_i(z;\b)$ by
\begin{eqnarray}
h_i(z;\beta)=-\sum_{n \ne 0}{h^i_n \over
[n]}q^{-\beta|n|}z^{-n} +Q_{h^i}+h^i_0\ln z,
\end{eqnarray}
where $\beta$ is a parameter.
Other bosonic fields are defined similarly.

The Drinfeld generators at level-one are realized by the free boson fields as
\cite{Kim97}
\begin{eqnarray*}
& & h^1_m = a^1_mq^{-|m|/2}-a^{2}_mq^{|m|/2},
~h^{2}_m =a^{2}_mq^{-|m|/2}+b_mq^{-|m|/2},~~~m\in{\bf Z},\\
& &X^{\pm}_1(z) = \pm :e^{\pm h_1(z;\pm\frac{1}{2})}:e^{\pm i\pi a^1_0},
~X^{+}_2(z) = :e^{h_{2}(z;\frac{1}{2})}e^{c(z;0)}: e^{-i\pi a^1_0},\\
& &X^{-}_2(z)= :e^{-h_{2}(z;-\frac{1}{2})}[ \partial_z
e^{-c(z;0)}]:e^{i\pi a^1_0},~\gamma=q,
\end{eqnarray*}
where  $\partial_z f(z)={f(qz)-f(q^{-1}z) \over (q-q^{-1})z}$
$:O:$ stands for the usual normal ordering of $O$.

Consider the bosonic Fock spaces 
$ F_{\lambda_1,\lambda_2,\lambda_3,\lambda_4}$,
generated by $a^{i}_{-m},b_{-m},c_{-m}$ $(m>0)$ over the vacuum
vectors $|\lambda_1,\lambda_2,\lambda_3,\lambda_4>$,
\begin{eqnarray}
 F_{\lambda_1,\lambda_2,\lambda_3,\lambda_4}={\bf
C}[a^{i}_{-1},a^{i}_{-2},...
;b_{-1},..;c_{-1},..]|\lambda_1,\lambda_2,\lambda_3,\lambda_4>,
\end{eqnarray}
 where
\begin{eqnarray*}
&&a^{i}_m|\lambda_1,\lambda_2,\lambda_3,\lambda_4>=0,~~
b_m|\lambda_1,\lambda_2,\lambda_3,\lambda_4>=0,~~
c_m|\lambda_1,\lambda_2,\lambda_3,\lambda_4>=0,~~{\rm for }\ \ m>0,\\
&&|\lambda_1,\lambda_2,\lambda_3,\lambda_4>=e^{\lambda_1Q_{a^1}+
\lambda_2Q_{a^2}+\lambda_3Q_{b}+\lambda_4Q_c}|0,0,0,0>.
\end{eqnarray*}
Introduce the following spaces
\begin{eqnarray}
F_{(\alpha;\beta)}=\bigoplus_{i,j\in {\bf Z}} F
_{\beta+i,\beta-i+j,\beta-\alpha+j,-\alpha+j}.
\end{eqnarray}
It can be shown that the bosonized action of $\uqs2h$ on $F_{(\a;\b)}$
is closed.
To obtain the irreducible subspaces in $F_{(\alpha;\beta)}$, it convenient
to introduce a pair of fermionic currents\cite{Bou89,Kim97}
\begin{eqnarray*}
\eta(z)=\sum_{n\in{\bf Z}}\eta_nz^{-n-1}=:e^{c(z;0)}:,\ \
\xi(z)=\sum_{n\in{\bf Z}}\xi_nz^{-n}=:e^{-c(z;0)}:,
\end{eqnarray*}
The mode expansion of $\eta(z),\xi(z)$ is well defined on $
F_{(\alpha;\beta)}$ for $\alpha\in{\bf Z}$, and it satisfies the following
relation
\begin{eqnarray*}
\xi_m\xi_n+\xi_n\xi_m=\eta_m\eta_n+\eta_n\eta_m=0,\ \
\xi_m\eta_n+\eta_n\xi_m=\delta_{m,n}.
\end{eqnarray*}
Since $\eta_0$ commutes (or anticommutes) with $\uqs2h$, $\eta_0$
plays the role of screening charge and $\eta_0\xi_0$ qualify as the projector
from $F_{(\a;\b)}$ to the Kernel of $\eta_0$.
Set $\l_{\a}=(1-\a)\L_0+\a\L_2,~\a\in {\bf
Z}$, where $\L_i(i=0,1,2)$ are the fundamental weights of
$\uqs2h$, and
\bea
\u_{\a}=\lt\{\begin{array}{ll}
    \L_{\a}, & \a=0,1,2 \\
    \l_{\a-1}& {\rm for}~\a>2\\
    \l_{\a}&{\rm for}~\a<0
\end{array}
\rt..\label{Wei}
\eea
Define $V(\u_{\a})=\eta_0\xi_0F_{(\a,\b-\a)}$. Following
\cite{Kim97,Yan99}, $V(\u_{\a})$ $(\a\in{\bf Z})$ are the
irreducible highest weight $\uqs2h$-modules with the
highest weight $\u_{\a}$.

\subsection{Level-one Vertex operators}

Let $V(\lambda)$ be the highest weight $\uqs2h$-module with the
highest weight $\lambda$. Consider the following intertwiners of
$\uqs2h$-modules:
\begin{eqnarray*}
& &\Phi_{\lambda}^{\mu V}(z) :
 V(\lambda) \longrightarrow V(\mu)\otimes V_{z} ,~~~~
\Phi_{\lambda}^{\mu V^{*}}(z) :
 V(\lambda) \longrightarrow V(\mu)\otimes V_{z}^{*S}.
\end{eqnarray*}
They are intertwiners in the sense that for any $x\in \uqs2h$,
\begin{eqnarray}
\label{EX}
\Theta(z)\cdot x=\Delta(x)\cdot \Theta(z),\ \ \Theta(z)=
\Phi(z),\Phi^{*}(z),
\end{eqnarray}
the grading of these operators is :$[\Theta(z)]=0$. $\Phi(z)$
is called type I (dual) vertex operator \cite{Jim94}.
We expend the vertex operator as
\begin{eqnarray}
& &\Phi(z)=\sum_{j=1,2,3}\Phi(z)_j\otimes v_j\ \ ,\ \
\Phi^{*}(z)=\sum_{j=1,2,3}\Phi^{*}(z)_j\otimes
v^{*S}_j.\label{VoxI}
\end{eqnarray}
Define the operators $\phi_j(z),\phi^{*}_j(z),\psi_j(z)$ and
$\psi^{*}_j(z)$ $(j=1,2,3)$ by
\bea
& &\phi_3(z)=:e^{-h^{*}_2(q^2z;-\frac{1}{2})+c(q^2z;0)}:
e^{-i\pi  a^{2}_{0}},\label{VOX1}\\
& & \phi_2(z)=-[\phi_3(z),X_0^{-,2}]_{q^{-1}},~
\phi_1(z)=[\phi_2(z),X_0^{-,1}]_q,\\
& &\phi^{*}_1(z)=:e^{h^{*}_1(qz;-\frac{1}{2})}:e^{i\pi a^{2}_{0}},\\
& &\phi^{*}_2(z)=-q^{-1}[\phi^{*}_1(z),X_0^{-,1}]_{q},~
\phi^{*}_3(z)=q^{-1}[\phi^{*}_2(z),X_0^{-,2}]_q,\label{VOX2}
\eea
where $h^{*1}_{m}=-h^2_m$,
$h^{*2}_m=-h^1_m-\frac{[2m]}{[m]}h^2_m$ and $Q_{h^{*1}}=-Q_{h^2}$,
$ Q_{h^{*2}}=-Q_{h^1}-2Q_{h^2}$. Since the operator
$\phi_i(z)$, $\phi^*_i(z)$ commute (or anti-commute) with
$\eta_0$, we define
\bea
\Phi_i(z)=\eta_0\xi_0\phi_i(z)\eta_0\xi_0,~~~~~
\Phi^*_i(z)=\eta_0\xi_0\phi^*_i(z)\eta_0\xi_0.\label{VoxI1}
\eea
According \cite{Kim97,Yan99}, the vertex operators $\Phi(z)$ and $\Phi^*(z)$
 (\ref{VoxI}) given by (\ref{VoxI1}) are the only 
 type I vertex operators of $\uqs2h$
 which intertwine the level-one irreducible highest weight $\uqs2h$-modules
 $V(\u_{\a})$ $(\a\in {\bf Z})$
\bea
\Phi(z):~V(\u_{\a})\longrightarrow V(\u_{\a-1})\otimes V_z,
~~\Phi^*(z):~V(\u_{\a})\longrightarrow V(\u_{\a+1})\otimes
V^{*s}_z.\no
\eea
It is shown \cite{Yan99} that the above vertex operators
satisfy the graded Faddeev-Zamolodchikov algebra
\bea
& &\Phi_j(z_2)\Phi_i(z_1)=\sum_{kl}R(\frac{z_1}{z_2})^{kl}_{ij}
\Phi_k(z_1)\Phi_l(z_2)(-1)^{[i][j]},\label{ZF1}\\
& &\Phi^{*}_j(z_2)\Phi^{*}_i(z_1)=\sum_{kl}R(\frac{z_1}{z_2})^{ij}_{kl}
\Phi^{*}_k(z_1)\Phi^{*}_l(z_2)(-1)^{[i][j]},\label{ZF2}\\
& &\Phi_j(z_2)\Phi^{*}_i(z_1)=\sum_{kl}\bar{R}(\frac{z_1}{z_2})_{ij}^{kl}
\Phi^{*}_k(z_1)\Phi_l(z_2)(-1)^{[k][l]},\label{ZF3}
\eea
where $\bar{R}(z)=R^{-1,st_1}(z)$.
Moreover, the vertex operators having the following
invertibility relations
\bea
& &\Phi_i(z)\Phi^{*}_j|_{V(\Lambda_{\alpha})}
=-(-1)^{[j]}\delta_{ij}id_{V(\Lambda_{\alpha})} ,\label{DEF1}\\
& &-\sum_{k}(-1)^{[k]}\Phi^{*}_k(z)\Phi_k(z)
|_{V(\Lambda_{\alpha})}=id|_{V(\Lambda_{\alpha)}} ,\nonumber\\
& &\Phi^{*}_i(zq^2)\Phi_j(z)|_{V(\Lambda_{\alpha})}
=\delta_{ij}q^{2\rho_i}id|_{V(\Lambda_{\alpha})},\label{DEF2}\\
& &\sum_{k}q^{-2\rho_k}\Phi_k(z)\Phi^{*}_k(zq^2)
|_{V(\Lambda_{\alpha})}=id|_{V(\Lambda_{\alpha})} .\nonumber
\eea

\sect{Appendix B}
In this appendix, we give the normal ordering relations of fundamental bosonic
fields:
\begin{eqnarray*}
& &e^{h_1(z_1;\beta_1)}e^{h_1(z_2;\beta_2)}=(z_1-q^{-(\beta_1+\beta_2)+1}z_2)
(z_1-q^{-(\beta_1+\beta_2)-1}z_2):e^{h_1(z_1;\beta_1)}e^{h_1(z_2;\beta_2)}:,\\
& &e^{h_1(z_1;\beta_1)}e^{h_2(z_2;\beta_2)}=\frac{1}
{z_1-q^{-(\beta_1+\beta_2)}z_2}:e^{h_1(z_1;\beta_1)}e^{h_2(z_2;\beta_2)}:,\\
& &e^{h_2(z_1;\beta_1)}e^{h_1(z_2;\beta_2)}=\frac{1}
{z_1-q^{-(\beta_1+\beta_2)}z_2}:e^{h_2(z_1;\beta_1)}e^{h_1(z_2;\beta_2)}:,\\
& &e^{h_2(z_1;\beta_1)}e^{h_2(z_2;\beta_2)}=
:e^{h_2(z_1;\beta_1)}e^{h_2(z_2;\beta_2)}:,\\
& &e^{h_i(z_1;\beta_1)}e^{h^{*}_j(z_2;\beta_2)}=
(z_1-q^{-(\beta_1+\beta_2)}z_2)^{\delta_{ij}}:
e^{h_i(z_1;\beta_1)}e^{h^{*}_j(z_2;\beta_2)}:,\\
& &e^{h^{*}_i(z_1;\beta_1)}e^{h_j(z_2;\beta_2)}=
(z_1-q^{-(\beta_1+\beta_2)}z_2)^{\delta_{ij}}:
e^{h^{*}_i(z_1;\beta_1)}e^{h_j(z_2;\beta_2)}:,\\
& &e^{h^{*}_1(z_1;\beta_1)}e^{h^{*}_1(z_2;\beta_2)}=
:e^{h^{*}_1(z_1;\beta_1)}e^{h^{*}_1(z_2;\beta_2)}:,\\
& &e^{h^{*}_1(z_1;\beta_1)}e^{h^{*}_2(z_2;\beta_2)}=\frac{1}
{z_1-q^{-(\beta_1+\beta_2)}z_2}:e^{h^{*}_1(z_1;\beta_1)}
e^{h^{*}_2(z_2;\beta_2)}:,\\
& &e^{h^{*}_2(z_1;\beta_1)}e^{h^{*}_1(z_2;\beta_2)}=\frac{1}
{z_1-q^{-(\beta_1+\beta_2)}z_2}:e^{h^{*}_2(z_1;\beta_1)}
e^{h^{*}_1(z_2;\beta_2)}:,\\
& &e^{h^{*}_2(z_1;\beta_1)}e^{h^{*}_2(z_2;\beta_2)}=\frac{1}
{(z_1-q^{-(\beta_1+\beta_2)+1}z_2)(z_1-q^{-(\beta_1+\beta_2)-1}z_2)}
:e^{h^{*}_2(z_1;\beta_1)}e^{h^{*}_2(z_2;\beta_2)}:,\\
& &e^{c(z_1;\beta_1)}e^{c(z_2;\beta_2)}=(z_1-q^{-(\beta_1+\beta_2)}z_2)
:e^{c(z_1;\beta_1)}e^{c(z_2;\beta_2)}:.
\end{eqnarray*}

\sect{Appendix C}
We here summarize the formulas concerning coherent states of bosons which
have been used in section 5.

The coherent states $|\zeta^1,\zeta^2,\zeta^3>$ and
$<\bar{\zeta}^1,\bar{\zeta}^2,\bar{\zeta}^3|$
in the Fock space $F_{(0;\b)}$
and its dual space $F^*_{(0;\b)}$ are defined by
\bea
&&|\zeta^1,\zeta^2,\zeta^3>=exp\lt\{\sum_{m=1}\sum_{i=1}^{2}\frac{m}{[m]^2}
\zeta^i_mh^{*i}_{-m}+\sum_{m=1}\frac{m}{[m]^2}\zeta^3_{m}c_{-m}\rt\}
|\b,\b,\b,0>,\\
&&<\bar{\zeta}^1,\bar{\zeta}^2,\bar{\zeta}^3|=<\b,\b,\b,0|exp\lt\{
\sum_{m=1}\sum_{i=1}^{2}\frac{m}{[m]^2}
\bar{\zeta}^i_mh^{*i}_m+\sum_{m=1}\frac{m}{[m]^2}\bar{\zeta}
^3c_m\rt\}
\eea
where $\zeta_m^l$ and ${\bar\zeta}^l_m$ $(l=1,2,3,~m=1,2,\cdots)$ are
complex conjugate parameters.

Noting that
\bea
&&h^i_m |\b,\b,\b,0>=0,~~<\b,\b,\b,0|h^i_{-m}=0,~i=1,2, ~~~m\geq1,\no\\
&&c_m |\b,\b,\b,0>=0,~~<\b,\b,\b,0|c_{-m}=0, ~~~m\geq1,\no
\eea
one can easily verify
\bea
&&h^i_m|\zeta^1,\zeta^2,\zeta^3>=\zeta^i_m|\zeta^1,\zeta^2,\zeta^3>,~~
<\bar{\zeta}^1,\bar{\zeta}^2,\bar{\zeta}^3|h^i_{-m}=
\bar{\zeta}^i_m<\bar{\zeta}^1,\bar{\zeta}^2,\bar{\zeta}^3|,~~i=1,2,\no\\
&&c_m|\zeta^1,\zeta^2,\zeta^3>=\zeta^3_m|\zeta^1,\zeta^2,\zeta^3>,~~
<\bar{\zeta}^1,\bar{\zeta}^2,\bar{\zeta}^3|c_{-m}=
\bar{\zeta}^3_m<\bar{\zeta}^1,\bar{\zeta}^2,\bar{\zeta}^3|.\no
\eea
One can also show that the coherent states 
$\{|\zeta^1,\zeta^2,\zeta^3>\}$ (resp.
$<\bar{\zeta}^1,\bar{\zeta}^2,\bar{\zeta}^3|\}$) 
form a complete basis in Fock space
$F_{(0;\b)})$ (resp. $F_{(0;\b)}^*$). Namely, one can
verify the completeness relation
\bea
id_{F_{(0;\b)}}&=&
\int \prod_{m=1}^{\infty}~
\frac{d\zeta^1_md\bar{\zeta}^1_m~d\zeta^2_md\bar{\zeta}^2_m
~d\zeta^3_md\bar{\zeta}^3_m}
{\frac{[m]^2}{m}~det(\frac{[a_{ij}m][m]}{m})}
exp\lt\{-\sum_{m=1}^{\infty}
\sum_{i,j=1}^2~\frac{K_{ij}(m)m}
{[m]}\zeta^i_m\bar{\zeta}^j_m
\rt\}\no\\
&&~~~~~\times|\zeta^1,\zeta^2,\zeta^3>
<\bar{\zeta}^1,\bar{\zeta}^2,\bar{\zeta}^3|,
\label{Coh}
\eea
where $K_{ij}(n)$ is a $2\times 2$ matrix satisfying
\bea
\sum_{l=1}^{2}K_{il}(n)[a_{lj}n]=\d_{ij}.\no
\eea
One may also derive the following identity
\bea
&&\int \prod_{m=1}^{\infty}~
\frac{d\zeta^1_md\bar{\zeta}^1_m~d\zeta^2_md\bar{\zeta}^2_m
~d\zeta^3_md\bar{\zeta}^3_m}
{\frac{[m]^2}{m}~det(\frac{[a_{ij}m][m]}{m})}
exp\lt\{
-\frac{1}{2}\sum_{m=1}^{\infty}\l_m~\lt(\bar{\zeta}^1_m,
\bar{\zeta}^2_m,\bar{\zeta}^3_m,
\zeta^1_m,\zeta^2_m,\zeta^3_m\rt){\cal A}_m\lt(
\begin{array}{l}\bar{\zeta}^1_m\\
\bar{\zeta}^2_m\\
\bar{\zeta}^3_m\\
\zeta^1_m\\
\zeta^2_m\\
\zeta^3_m
\end{array}\rt)\rt.\no\\
&&~~~~~~\lt.+\sum_{m=1}^{\infty}
(\bar{\zeta}^1_m,\bar{\zeta}^2_m,\bar{\zeta}^3_m,
\zeta^1_m,\zeta^2_m,\zeta^3_m){\cal B}_m\rt\}\no\\
&&~~=\prod_{m=1}^{\infty}(-det{\cal A}_m)^{-\frac{1}{2}}exp\lt\{
\frac{1}{2}\sum_{m=1}^{\infty}
\frac{[m]^2}{m}~det\lt( \frac{[a_{ij}m][m]}{m}\rt)
{\cal B}_m^t{\cal A}_m^{-1}
{\cal B}_m\rt\},\label{ID}
\eea
where ${\cal A}_m$ are invertible constant $6\times 6$ matrices and
${\cal B}_m$ are constant $6$ component vectors.

\bibliographystyle{unsrt}

\begin{thebibliography}{10}
\bibitem{Ess92} F.H.L. Essler, V.E. Korepin, K. Schoutens,
{\it Phys. Rev. Lett.} {\bf 68} (1992), 2960.
\bibitem{Foe93} A. Foerster, M. Karowski, {\it Nucl. Phys.
} {\bf B396} (1993), 611.
\bibitem{Bra95} A.J. Bracken, M.D. Gould, J.R. Links, Y.-Z.
Zhang, {\it Phys. Rev. Lett} {\bf 74} (1995), 2768.
\bibitem{Pfa96} M.P. Pfannmuller, H. Frahm, {\it Nucl.
Phys. } {\bf B479} (1996), 575.
\bibitem{Ram96} P.B. Ramos, M.J. Martins, {\it Nucl. Phys.}
 {\bf B479} (1996), 678.
\bibitem{Fan00} H. Fan, M. Wadati, X.M. Wang, {\it Phys.
Rev. }{\bf B61}(2000), 3450.
\bibitem{Yan99} W.-L. Yang, Y.-Z. Zhang, {\it Nucl. Phys. }{\bf B547}
(1999), 599; {\it Jour. Math. Phys.} {\bf 41} (2000), 5849.
\bibitem{Dav93} B. Davies, O. Foda, M. Jimbo, T. Miwa, A. Nakayashiki,
{\it  Commun. Math. Phys.} {\bf 151}(1993) , 89.
\bibitem{Jim94} M. Jimbo, T. Miwa, {\it Algebraic analysis of solvable
lattice model}, CBMS Regional Conference Series in Mathematics,
{\bf Vol. 85} (AMS, Providence, 1994).
\bibitem{Fre92} I.B. Frenkel, N.Yu. Reshetikhin, {\it Commmun. Math.
Phys.} {\bf 146} (1992), 1.
\bibitem{Idz94} M.Idzumi, {\it Int. J. Mod. Phys.} {\bf A9} (1994),
449; H.Bougurzi and R.Weston , {\it Nucl. Phys.} {B417} (1994),
439.
\bibitem{Koy94} Y.Koyama, {\it Commun. Math. Phys.} {\bf 164}(1994), 277.
\bibitem{Luk96} S.Lukyanov and Y.Pugai, {\it Nucl. Phys. } {\bf
B473}(1996), 631;
Y.Asai, M.Jimbo, T.Miwa and Y.Pugai, {\it Jour. Phys.} {\bf A29} (1996),
6595.
\bibitem{Hon98} J.Hong, S.J.Kang, T.Miwa and R.Weston, {\it J. Phys. }
{\bf A31} (1998), L515.
\bibitem{Hou99} B.Y. Hou, W.-L. Yang, Y.-Z. Zhang, {\it Nucl. Phys. }
{\bf B556} (1999), 485.
\bibitem{Jim95} M. Jimbo, R. Kedem, T. Kojima, H. Konno, T. Miwa,
{\it Nucl. Phys. }{\bf B441} (1995), 437.
\bibitem{Sky88} E.K. Sklyanin, {\it Jour. Phys. }{ \bf A21} (1988), 2375.
\bibitem{Gho94} S. Ghoshal, A. Zamolodchikov,
{\it Int. J. Mod. Phys. A} {\bf 21}(1994), 3841.
\bibitem{Miw97} T. Miwa, R. Weston, {\it Nucl. Phys.} {\bf B} (1997), 517;
B.Y. Hou, W.-L. Yang, {\it Commun. Theor. Phys.} {\bf 27 }(1997), 257.
\bibitem{Fur00} H. Furutsu, T. Kojima, e-print {\it solv-int/9905009};
T. Kojima, Y. -H. Quano, e-print {\it nlin.SI/0001038}.
\bibitem{Yan01} W.-L. Yang, Y.-Z. Zhang, Izergin-Korepin
model with a boundary, {\it Nucl. Phys.} {\bf B} (2001), in
press.
\bibitem{Hou97} B. Y. Hou, K.J. Shi, Y.S. Wang, W.-L. Yang, {\it Jour.
Phys. }{\bf A30}(1997), 251; {\it
Int. Jour. Mod. Phys.} {\bf A12} (1997), 1711; H. Furutsu, T. Kojima,
Y.-H. Quano, e-print {\it solv-int/9910012}.
\bibitem{Doi99} B.Y. Hou, K. J. Shi, W.-L. Yang, {\it Commun. Theor.
Phys.} {\bf 31} (1999), 265; A.Doikou, R. Nepomechie,
{\it Phys. Lett.} {\bf B462} (1999), 121.
\bibitem{Bra98} A.J. Bracken, X.Y. Ge, Y.-Z. Zhang, H.Q.
Zhou, {\it Nucl. Phys. }{\bf B516} (1998), 588.
\bibitem{Kim97} K. Kimura, J. Shiraishi, J. Uchiyama, {\it
Commun. Math. Phys. } {\bf 188} (1997), 367.
\bibitem{Bou89} P. Bouwknegt, A. Ceresole, J. G. McCarthy,
P. van Nieuwenhuizen, {\it Phys. Rev. } {\bf D39} (1989),
2971.
\end{thebibliography}

\end{document}